\documentclass[conference]{IEEEtran}
\ifCLASSINFOpdf
\usepackage[pdftex]{graphicx}
\usepackage{xcolor}
\usepackage{url}

  \graphicspath{{./pdf/}{../jpeg/}}
\else
\fi
\begin{document}

\title{Correlating power consumption and network traffic for improving data centers resiliency}

\author{\IEEEauthorblockN{Roberto Baldoni \IEEEauthorrefmark{1},
Mario Caruso  \IEEEauthorrefmark{1}
Adriano Cerocchi \IEEEauthorrefmark{2}
Claudio Ciccotelli \IEEEauthorrefmark{1}
Luca Montanari \IEEEauthorrefmark{1}}
Luca Nicoletti \IEEEauthorrefmark{3}

\\
\IEEEauthorblockA{\IEEEauthorrefmark{1}Cyber Intelligence and Information Security Research Center, ``Sapienza'' University of Rome, Italy }\IEEEauthorblockA{\IEEEauthorrefmark{3} SOGEI, Via M. Carucci n. 99, Rome, Italy}
\IEEEauthorblockA{\IEEEauthorrefmark{2} Over Technologies, Via G. Peroni 442, Rome, Italy\\}}

%


\maketitle

\begin{abstract}

The deployment of business critical applications and information infrastructures are moving to the cloud. This means they are hosted in large scale data centers with other business applications and infrastructures  with less (or none) mission critical constraints. This mixed and complex environment makes very challenging the process of monitoring critical applications and handling (detecting and recovering) possible failures of servers' data center that could affect responsiveness and/or reliability of mission critical applications.  
Monitoring mechanisms used in data center are usually intrusive in the sense that they need to install agents on each single server. This has considerable drawbacks: huge usage of human resources to install and patch the system and interference with the critical application because agents share application resources. 

In order to detect (and possibly predict) failures in data centers the paper does a first attempt in showing the correlation between network traffic and servers' power consumption.   
This is an important step in deriving non-intrusive monitoring systems, as both network traffic and power consumption can be captured without installing any software at the servers. This will improve in its turn the overall resiliency of the data center and its self-managing capacity.





\end{abstract}


%
\IEEEpeerreviewmaketitle

\section{Introduction}


Consider as examples very diverse applications such as a large scale payroll for the entire public administration of a country, consumer credit support or a system running a securities market. These are mission critical applications for the stability of the country. Thus they need to run with very strict non-functional requirements in terms of reliability, security and responsiveness to application end-users. For economic reasons, it is expected that such applications will migrate on a ``public administration'' private cloud shortly, where they will coexist with other applications that have less stringent (or none) non-functional requirements. Such cloud will reside on as few large data centers as possible in order to get the maximum saving and, at the same time, to concentrate the IT control. Each of these data centers will include thousands of servers, large scale storage and communication infrastructure. Thus a key role for satisfying requirements of a mission critical application is carried out by a monitoring system that is able to report and to timely react to failures happening in the data center. Such failures could indeed either deteriorate the performance or block a mission critical application. 

Thus this paper concentrates on the design of a monitoring system based on the information coming from network traffic exchanged among servers and servers power consumption. The aim of this paper is to show the existence of correlation between network traffic and power consumption that can be used to asses (and to possibly predict) failures in the data centers.
To highlight this correlation we did an extensive experimentation in one of the four data centers of the 
Italian Ministry of Economy and Finance (MEF).

MEF is the executive body responsible for economic, financial and budget policy, planning of public investment, coordinating public
expenditure and verifying its trends, revenue policies and tax system in Italy. The entire MEF infrastructure is designed to have a high resiliency degree, to prevent the complete interruption of a service in case of failure. The MEF IT ecosystem is always under the control of an advanced and complex monitoring system, which continuously checks the health state of hardware and software, of the network and of the end-user experience.
Nonetheless, monitoring systems employed at MEF have two main  point of weakness: (i) the approach is reactive: able only to react to failure events when they have already occurred and (ii) monitoring systems require the installation on each server of software probes, determining a huge impact in terms o management effort and procedures. 


Thus, in the design of a new monitoring system, we highlighted the following two properties:
(i) no deployment of software probes (\textit{non-intrusive} monitoring), and (ii) agnostic with respect to applications running in the data center, (\textit{black-box} monitoring). 
Non-intrusive monitoring can be realized observing network traffic and power consumption, without deploying any software probes, but relying on sniffing network packets (without introducing packets or delay in the network) and reading power consumption through smart Power Distribution Unit (smart-PDU) in order to accurately measure the power consumption at each rack enclosure. After several months of experiments we were able to demonstrate that there is correlation between network traffic and power consumption and this correlation can be used to design failure prediction techniques, to improve data centers resiliency. This phase of experimentation is reported in the paper.

The rest of the paper is organized as follows. Section \ref{sec:relwork} presents the related work; Section \ref{sec:arch} describes the major issues in data center monitoring and introduces an architecture for this; Section \ref{sec:exper} describes the first experimental campaign and describe the MEF data center that we used as testbed; Section \ref{sec:concl} concludes the paper.

\section{Related Work} \label{sec:relwork}

Monitoring based only on network traffic is recognized to be non-intrusive and black-box, meaning that (i) no application-level knowledge is needed to perform the monitoring \cite{Narasimhan,Aguilera,BaldoniLMMR12}, and (ii) the monitor mechanism does not install software on the monitored system \cite{BaldoniLMMR12}. In \cite{BaldoniLMMR12} CASPER is presented, a non-intrusive and black-box approach to monitor air traffic control systems. It uses network traffic only in order to represent the system health so as to recognize deviations thus triggering failure prediction. At the best of our knowledge, this is the only work that is both non-intrusive and black-box. In the context of symptoms monitoring mechanisms, there exist research works that use black-box approaches, i.e., no knowledge of the applications of the system is required. Narasimhan et. all \cite{Narasimhan} introduce Tiresias, a black-box failure prediction system that considers symptoms generated by faults in distributed systems. Tiresias uses a set of performance metrics that are system-level metrics, (e.g., UNIX proc file system metrics) and network traffic metrics. Tan et al.~\cite{TanGu} presents ALERT an anomaly prediction system that considers the hosts of the monitored system as black-boxes. Specifically, it uses sensors deployed in all the hosts of the controlled infrastructure to continuously monitor a set of metrics concerning CPU consumption, memory usage, input/output data rate. In this sense, ALERT can be categorized as an intrusive monitoring system that makes use of triple-state decision trees in order to classify component states of the monitored system. The authors in \cite{Aguilera} consider the problem of discovering performance bottlenecks in large scale distributed systems consisting of black-box software components (usually without source code available and not accessible due to vendors restrictions). 
The system introduced in \cite{Aguilera} solves the problem by using message-level traces related to the activity of the monitored system in a non-intrusive fashion (passively and without any knowledge of node internals or semantics of messages).

Most of the studies on power consumption monitoring in data centers have been conducted in the context of power management and energy efficiency \cite{lefurgy2008,xiaorui2008,xiaorui2009}.
None of these works, however, concerns data centers resiliency. In \cite{heller2010} and \cite{xiaodong2012} network traffic is monitored with the aim of consolidating traffic flows onto a small set of links and switches so as to shut down unused network elements, thereby reducing power consumption.
However, there is no attempt to correlate network traffic and power consumption.
Methods for detection of cyber attacks that exploit monitored power consumption data have been investigated in the context of power network systems \cite{kosut2010,teixeira2010,hashimoto2011}. 
In \cite{hashimoto2011} local information about power consumption, generation, and power flow are used for distributed detection of cyber attacks in power networks. 
In \cite{kazandjieva2009} a study on correlation between power consumption data and utilization statistics (CPU load and network traffic) is presented. 
This work shows a strong correlation between power consumption and CPU load of desktop computers. In the context of server racks, the authors conclude that CPU utilization alone does not completely explain the variation of power consumption, suggesting that additional metrics should be taken into account.
Our work is considerably different for two reasons. First of all our study addresses resiliency, while \cite{kazandjieva2009} concerns energy efficiency.
Moreover, our work is based on a non-intrusive and black-box approach, while the monitoring infrastructure used in \cite{kazandjieva2009} is not black-box, nor non-intrusive. Indeed, it requires software installation on monitored machines to track CPU utilization and to collects network-traffic statistics through SNMP, thus introducing additional traffic in the network.

To the best of our knowledge no previous work has investigated the possibility to exploit the correlation between power consumption and network traffic to improve data centers resiliency.
\section{Architecture}\label{sec:arch}

\subsection{Critical Infrastructure data centers} \label{CIDC}

The data centers that manage IT services of critical infrastructures, such as ministries, financial institutions, transportations, power grid, and so on, are nowadays at a huge level of complexity. The requirements of fault tolerance and of resilience in case of failure are very stringent, thus implying an high level of replication, both hardware and software, active and passive. 
Hundreds of physical blade servers (rack-mount), each one hosting tens of virtual machines, are organized in rack enclosures (19-inch rack). Each enclosure can physically hosts up to about 45 rack units. A rack unit can be a blade server, a network device, such as a switch, or other kinds of devices, including cooling mechanisms. 
Data centers commonly used by public organizations, can be composed by hundreds of such enclosures, occupying big rooms, entire floors or multiple floors.
The power supply is ensured by at least two independent supply loops connected to the enclosure, with  overload protection systems. Power consumption is usually monitored using a centralized monitoring system.
From a network point of view, several switches connect servers within the single enclosure and among enclosures, according to policies of network segmentation, applied both for privacy and security reasons. 

All of the components in this environment are off-the-shelf and multiple vendors provide hardware in a single institution's data center. Heterogeneous and continuously changing hardware can also be found among the blade servers in a single enclosure, among the power supplies of multiple enclosures, among network devices of the same network and so on. 
All of this complexity and heterogeneity led to a situation in which the commonly used (and easiest) ways to take measurement about power consumption, or about network load, can have a significant deviation from the reality. This means that if one wants to monitor the actual network load of an enclosure, and the actual power consumption of it, the only way to do is by monitor the power distribution unit and the network switches of the enclosure. 

In the following, we describe how a dataset composed by actual network traces and actual power consumption traces of a single enclosure of a real critical infrastructure data center can be created. 

\subsection{Analyzing network traffic}
Several works have been presented in literature which aim is to monitor systems performance using also network data (i.e., packet rate, packet size, bandwidth etc) \cite{Narasimhan,Aguilera}. Some works use only network data to monitor system health \cite{BaldoniLMMR12}. These works have demonstrated that network can carry a large amount of data for monitoring purposes. Without examine the payload of packets, thus respecting privacy, sometimes network only can represent the level of health of a system.
An added value is given by the possibility of monitoring an existing system in a real \textit{non-intrusive} way: no software needs to be installed in servers, only a very small number of probes deployed properly can collect the information required. This has an extreme value for system administrators: a new way for monitoring and enhancing resiliency that requires only a small investment in time and money for the deployment. Note that, given the complexity described above of the modern data centers (\ref{CIDC}), installing a new monitor software requires a huge amount of human time or, if outsourced, a significant amount of economic resources. 
According to the network topology of the data center, network sniffers can be installed in order to monitor different granularity levels of network traffic. For example, if we imagine to monitor each single enclosure, network sniffers can be installed at the level of the switches of the enclosure to be monitored. This would provide performance indicators of all the hosts (typically virtual machines) running on that enclosure, e.g., for each host:
\begin{itemize}
\item inner/outer packet rate; 
\item inner/outer bandwidth; 
\item inner/outer message size.
\end{itemize}
Of course, same indicators can be computed for the whole enclosure, giving an overview of the traffic managed by the enclosure itself and of the load that it experiences in real time. 
The indicators produced can be used in real time in several ways, e.g., reported to a dashboard, used to classify the behavior of the system, used to instrument an inferential engine in order to recognize deviation from the correct behavior. These indicators gain an extreme value when correlated with other functional data for instance, power consumption. 

\subsection{Analyzing energy consumption} 


Usually data centers are equipped with energy monitors with the aim to (i) control the quality (picks analysis, harmonic distortion, etc) of the energy provided by the utility company and (ii) take care of the load balancing among the three phases of the three-phase energy provisioning.

In the recent years power consumption monitoring is getting finer granularity, sometimes providing energy consumed also by every single machine. Knowing this information is very useful: it allows user administrators to know which machines are really working. Power consumption contains a huge amount of information regarding CPU and storage usage, allowing the administrator to correctly balance the load among the machines.
Unfortunately, blade servers systems aggregate the consumption of all contained machines, thus hiding the power consumption of the single server. This problem can be solved by making use of very precise energy meters: a CPU performing a complex job always augment power consumption of the enclosure for a time that is similar to the task length. If that time results sufficiently long (e.g., a minute), it can be clearly identified in the consumption shape. Sometimes, a task fails into forcing its CPU in a loop state. Power consumption monitoring can reveal that very quickly, thus avoiding a domino effect that can affect several machines in the data center. 

Power consumption data, which monitoring can produce very relevant information to augment resiliency, are:
\begin{itemize}
	\item voltage;
	\item current;
	\item power factor.
\end{itemize}

Voltage and current can be used to compute power consumption shape, power factor is useful since can reveal in large advance an hardware failure: for example, if a power supply is loosing fast its own efficiency (i.e. a power factor running toward zero) it probably will encounter a failure. Having a power factor monitoring for each power supply 
can really help in this.

\subsection{Preliminary architecture for enhancing resiliency}
\begin{figure}[htbp] 
\begin{center}
\includegraphics[scale=0.33]{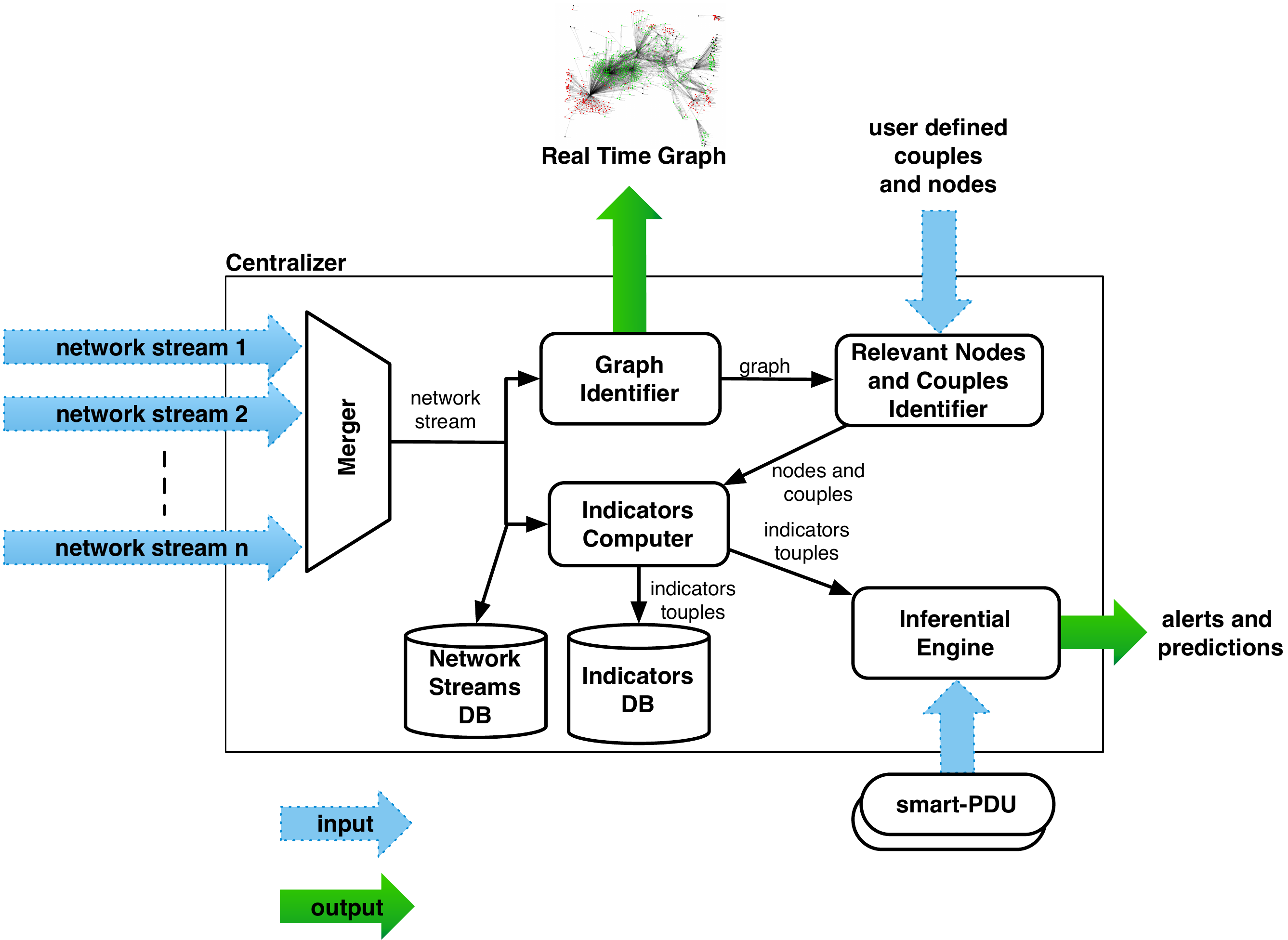}
\caption{Software Architecture of a non-intrusive black block monitoring system correlating network traffic and power consumption.}
\label{LLarch}
\end{center}
\end{figure}

Considering the advantages of having a non intrusive system to monitor and enhance resiliency of a critical infrastructure data center, we designed a preliminary architecture that correlates network data and power consumption in real time, for a single enclosure. 
The architecture is depicted in Figure \ref{LLarch}. The whole architecture lives inside a centralizer.  the centralizer can be an ordinary computer or a blade server. 
It takes in input (i) $n$ streams of network packets, directly produced by $n$ probes (network sniffers that capture packets from the switches of the enclosure); (ii) a stream of power consumption data from the smart-PDUs (that measure with high precision the power consumption of the monitored enclosure) and (iii) some user-defined data that we will explain below. The output of the architecture is (i) a Real Time Graph (RTG) reporting the topology of the network in real time along with several information about the nodes composing it with their interactions and (ii) alerts and prediction of failures, once the monitoring system recognized deviations from the correct enclosure behavior.
A description of all the modules composing the architecture is now provided. 

\paragraph{Merger} a software module that takes in input $n$ streams of captured packets and gives in output a single network stream opportunely merged\footnote{merging network traces is a solved problem, several tools are available. A synchronization of the probes is required e.g., a NTP server.}. 
\paragraph{Network Stream DB} Network traces can be stored for future analysis, for example in case of failure. Due to the big amount of data, network traces can be stored only for some days, according to the centralizer storage capability. The Network Stream DB maintains the past network traces for a fixed number of days. 
\paragraph{Graph Identifier} a software module that takes in input a stream of network packets and recognizes the topology of the network producing the stream. Complex statistics can be computed among the nodes interactions. The graph can be displayed in real time. 
\paragraph{Relevant Nodes and Couples Identifier (RNCI)} a software module that takes in input a graph from the Graph Identifier module and assesses which nodes or couples of nodes are particularly relevant. This module can also receive from a user known nodes identifiers (ip addresses) or couple of nodes which interactions are known to be relevant. The logic, according to which RNCI assesses nodes and couples, can use topology characteristics (e.g., fan in, fan out of nodes) or characteristics of the interaction among nodes (e.g., message rate between a couple, inner and outer message rate of a node, protocol used during interaction).
\paragraph{Indicators Computer} a software module that takes in input the network stream and, according to a set of rules, produces in real time indicators (e.g., message rate, bandwidth, message size, message rate per network protocol, message rate per physical machine). 
The indicators are grouped in tuples and produced in real time with a given frequency, for instance, one tuple per second. This led to have a snapshot of the observed system per second, for example, if we consider message rate, bandwidth, tcp messages, average message size, we would have a tuple, like the following, per second:
$$ <sec: 3; 4387 msg/s; 14042896 bps; 2632 tcp\_msgs; 400 byte>$$
meaning that during the third second of observation there have been 4387 messages, a mean bandwidth of 14042896 bit per second, 2632 tcp messages and an average message size of 400 bytes. 
Instead of producing a tuple representing the whole monitored system, the same tuple can be produced per single node or per couple of node, according to what RNCI module produces in output. 
In that case the tuple will contain also the ip address of the node (or the couple of ip addresses) which the tuple refers to. In the latter case, each second, Indicator Computer will produce a set of tuples, one representing the whole system and one per each node or couple provided by the RNCI module.
Indicators Computer module can be implemented combining Complex Event Processing techniques \cite{CEPBuchmann} and network statistics softwares (e.g., tcpstat~\cite{tcpstat}).
\paragraph{Indicators DB} Tuples of indicators can be stored for future analysis, for example in case of failure. Tuples are aggregated data, thus are an amount of data very reduced if compared with network traces. This means that tuples can be stored for a long period of time. Indicators DB maintains the past tuples of indicator for a number of months. 
\paragraph{Inferential Engine} a software module that using indicators tuples and power consumption data correlates them and, according to machine learning algorithms, triggers timely alerts or failure predictions if recognizes deviations from correct system behavior. 
The Inferential Engine is a crucial part of the architecture, that requires an accurate learning phase in order to build a knowledge base regarding the observed system.

\section{Experimental analysis}\label{sec:exper}
One of the pillars of this paper is the assumption that network data can be correlated with power consumption in order to enhance the resiliency of complex data centers. 
The scope of the first campaign of analysis has been answer to the following question: 
\begin{center}
\textit{Does it make sense correlating network data and power consumption in a real critical infrastructure data center?}
\end{center}
In order to provide an answer, we conducted a six months long experimental session along with Sogei s.p.a., a company of Italian Ministry of Economic and Finance (MEF) that manage the IT of the ministry. 
In particular we deployed part of the architecture presented before (section \ref{sec:arch}), in order to monitor a single enclosure of one of the data centers of MEF. Details about the data center follow. 
\subsection{Testbed: Italian Ministry of Economic and Finance data center} 
The MEF Data Center where we carried out experiments is a medium-size facility, featuring the following
main numbers: 80 physical servers; 250 virtual servers; 20 network devices; 8 security devices; more than 50 different Web Applications; 2 Storage Area Network with more than 6 TB of disk space; more than 1000 internal users and more 80.000 external managed single
users;
The network architecture has been conceived to conciliate security,
scalability and resiliency issues. It is divided into four main zones
with increasing levels of security; each zone hosts servers
and services with homogeneous requirements in terms of security and
privacy:
\begin{itemize}
\item The DMZ zone, to host all the systems used to allow the reachability
of MEF applications from the outside world;
\item The Web zone, to host the ``user/consumer logic'' layer of the applications; 
\item The Application zone, to host the ``business logic'' layer of the
applications;
\item The Database zone, to host data and DBMS.
\end{itemize}
All the network devices (switches, routers, firewalls) are new generation,
that simplifies the network fabric by converging both storage area
network (SAN) and LAN connectivity onto a single 10 Gigabit Ethernet
link. 
\subsection{Probes deployment}
\begin{figure}[htbp]
\begin{center}
\includegraphics[scale=0.4]{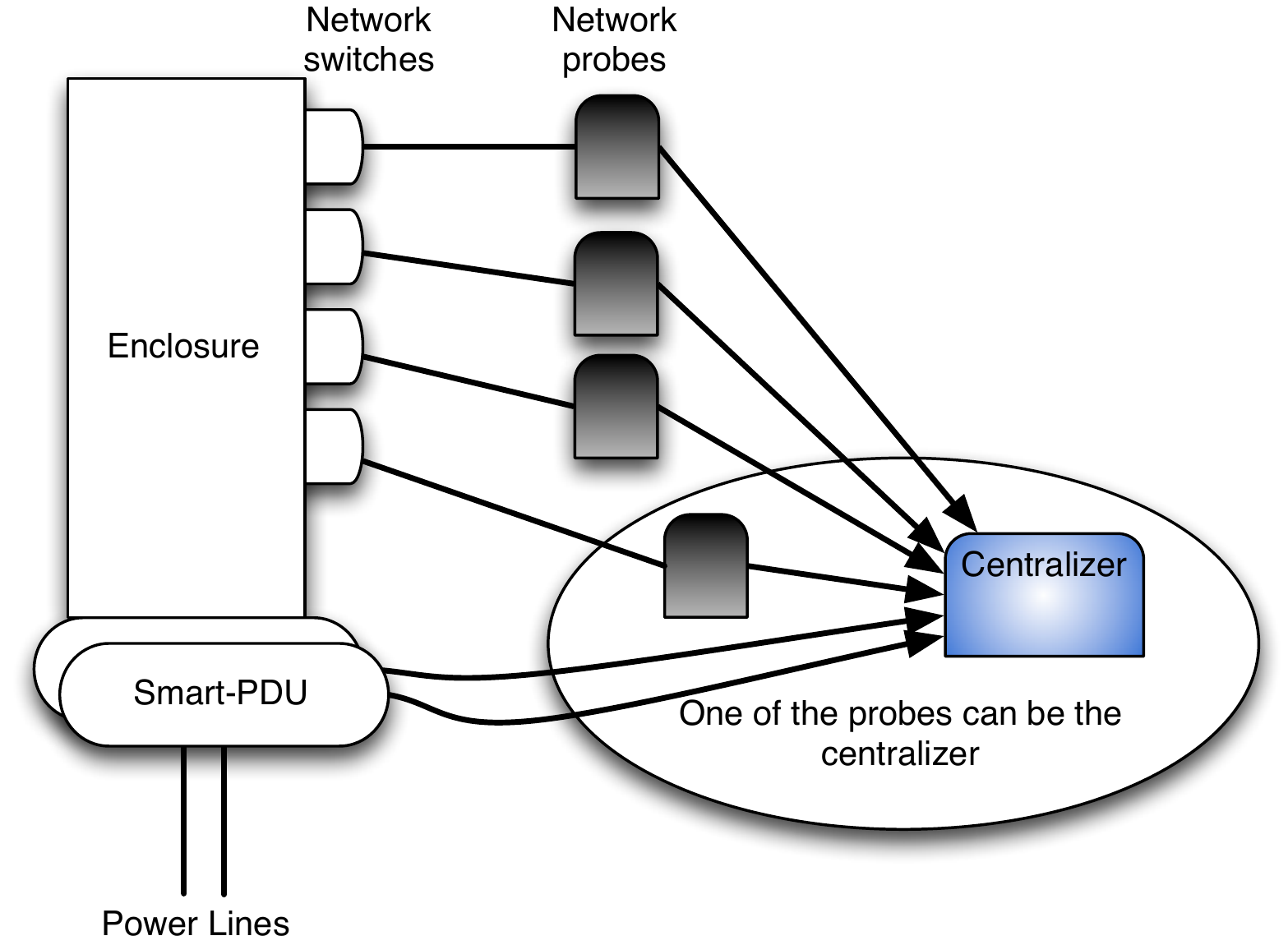}
\caption{Deployment of network probes and Smart-PDUs at  MEF data center.}
\label{HLarch}
\end{center}
\end{figure}
\begin{figure}[htbp]
\begin{center}
\includegraphics[scale=0.05]{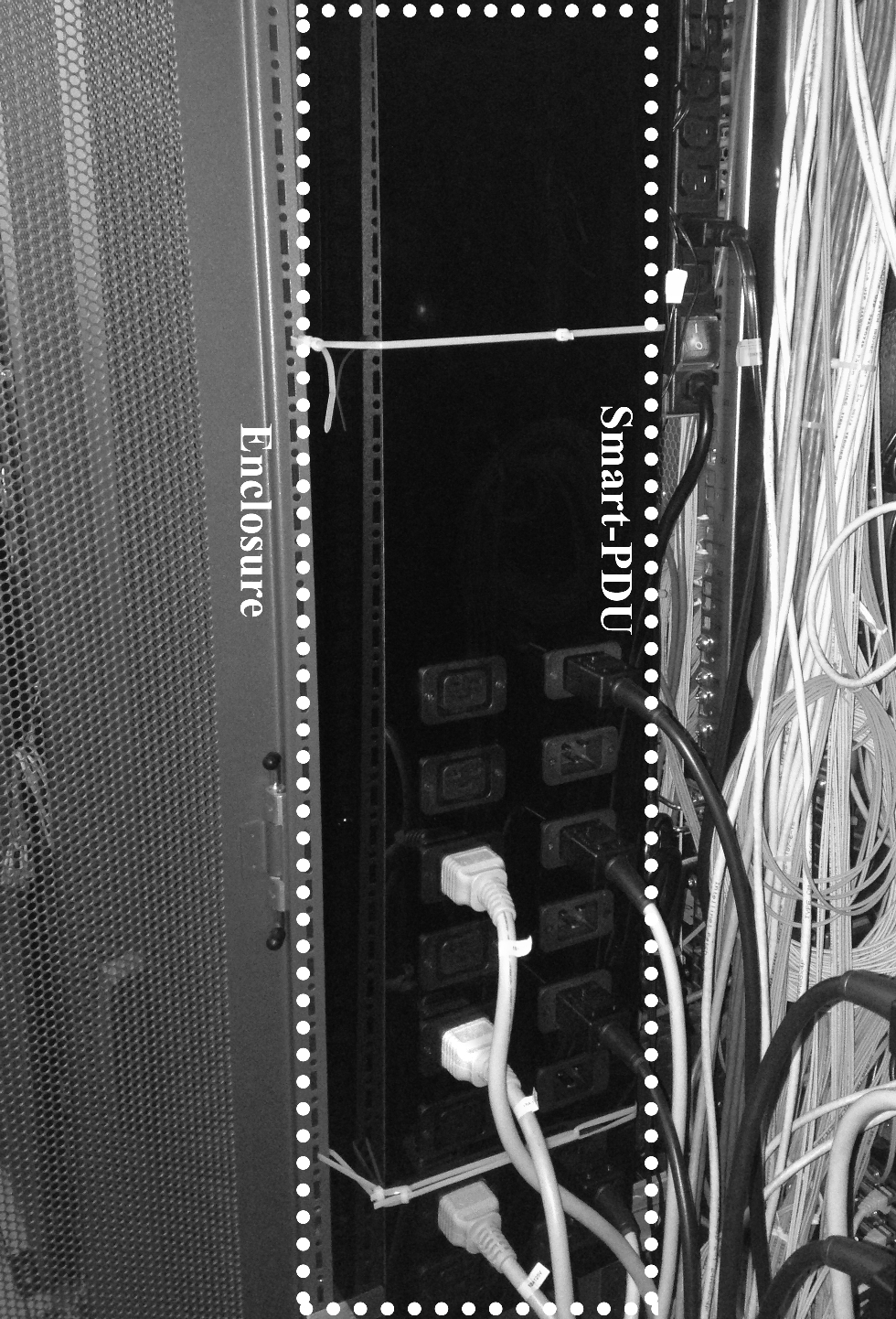}
\caption{Smart-PDUs deployement at MEF data center}
\label{foto}
\end{center}
\end{figure}

Figure \ref{HLarch} depicts how we deployed network probes and Smart-PDUs
in order to monitor one enclosure. The monitored enclosure uses four
network switches in order to allow its servers to communicate among
them and toward the rest of the data center. Since we want to monitor
all the traffic sent and received by the enclosure and all the traffic
among the servers inside the enclosure, we deployed four network sniffers
(four mini-pc Sapphire EDGE-HD3, with an additional USB ethernet adapter,
running tcpdump \cite{tcpdump}), one per network switch. One of the
Probes acts as centralizer, collecting data from the other three and
from two Smart-PDUs manufactured by Over \cite{over} that measure the power consumption of the entire
enclosure. The power consumption is sampled with a rate of one measure
every 10 seconds. Synchronization is achieved using an NTP server. An
additional network switch is required in order to grant communication
among probes and Smart-PDUs. All the deployed hardware is passive
with respect to the data center: it does not have an IP-address and
no network packets are introduced. Figure \ref{foto} is a picture
of one of the prototypal Smart-PDUs, behind the monitored enclosure.

\subsection{Dataset}
For this first part of the analysis, the probes collected network traces only. Correlation and study on captured data are done off-line. In this way, we built a dataset of approximately 2.5 Terabyte, representing the behavior of the monitored enclosure from a network and power consumption point of view, during the period 31 July 2013 - 31 January 2014.
The dataset is composed by \textit{pcap} network traces and a database of power consumption data\footnote{The DB of power consumption is in the format: $\langle$timestamp, active power, reactive power, phase displacement$\rangle$.}.

\subsection{Correlation}
Correlation between network traffic and power consumption would allow to make different hypothesis on the observed system. For example, if we assume the CPU as the main power consuming component, we could assess: 
\begin{enumerate}
\item a CPU intensive task can augment power consumption thus reducing network traffic (anticorrelation); 
\item a network intensive task can reduce power consumption due to reduced CPU load (anticorrelation);
\item a network and CPU idle period reduces network traffic and power consumption (correlation);
\item a network and CPU intensive task can augment power consumption (correlation).
\end{enumerate}
In order to verify the correlation among the data composing the dataset created, and to provide an answer to the question reported at the begin of this section, we graphed apparent power consumed by the enclosure, total packet rate (internal, toward and from the enclosure) and the \textit{population correlation coefficient} among them. 
The population correlation coefficient, also knows as \textit{Pearson product-moment correlation coefficient} $\rho_{X,Y}$, widely used as $corr(X,Y)$, is a measure of the linear correlation between two random variables $X$ and $Y$, and is defined as follows:
$$corr(X,Y) = \frac {cov(X,Y)}{\sigma_{X}\sigma_{Y}} = \frac {E[(X-\mu_{X})(Y-\mu_{Y})]}{\sigma_{X}\sigma_{Y}} $$
where $E$ is the expected value, $cov$ is the covariance, $\mu_{X}$, $\mu_{Y}$ is the expected value of $X$ and $Y$ respectively, $\sigma_{X}$ and $\sigma_{Y}$ are their standard deviations.
$corr(X,Y)$ returns a value between $-1$ and $1$. If $X$ and $Y$ are two independent variables, $corr(X,Y)=0$; If $|corr(X,Y)|>0.7$ there is a strong correlation, while if $0.3<|corr(X,Y)|<0.7$ the correlation is moderate. $corr(X,Y)>0$ in the case of a direct (increasing) linear relationship (correlation), $corr(X,Y)<0$ in the case of a decreasing (inverse) linear relationship (anticorrelation).

\begin{figure}[htbp]
\begin{center}
\includegraphics[scale=0.515]{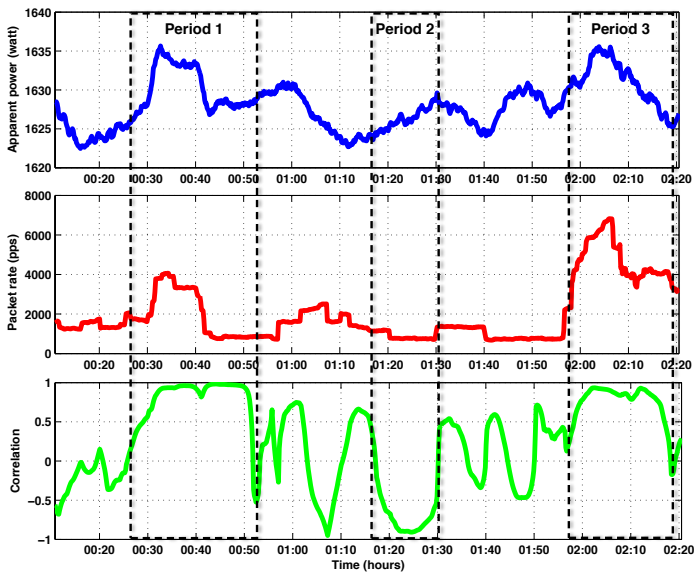}
\caption{Correlation between mean apparent power and mean packet rate.}
\label{Correl}
\end{center}
\end{figure}

Figure \ref{Correl} reports the obtained graphs for a two and a half hours long period of monitoring. The graph above reports the mean apparent power consumption of the enclosure, the graph in the center the mean packets rate, as defined before, and the graph below the correlation coefficient between packets per seconds and apparent power, during time. The mean is computed considering a sliding window of 10 minutes (see below). The temporal axes are aligned among the three graphs. 

During the period reported in Figure \ref{Correl} apparent power consumed by the enclosure is between 1622 W and 1636 W while packet rate is between 1000 pps and 7000 pps. 

In order to compute the correlation between power consumption and packet rate, we considered a 10 minutes sliding window, with a sampling rate of one sample every 10 seconds. In Figure \ref{Correl} we identified three periods in which the correlation has particularly interesting behavior. Period 1 is thirty minutes long and embodies a behavior in which the apparent power and packets per seconds increase and decrease in a very similar manner. This implies a very high level of correlation, near to 1, meaning that power consumption and network traffic are directly correlated. Period 3 highlights a behavior similar to Period 1. Period 2 highlights a period of anticorrelation: packets rate is decreasing while power consumption is increasing. 
According to the correlation definition, during the three periods we can see a strong correlation ($corr(X,Y)>|0.7|$). 
The absence of periods in which the correlation is 0 tells us that correlation between power consumption and network traffic exists. This has been observed in all the dataset produced during the experimentation. 

\section{Conclusions and future work}\label{sec:concl}

This work investigated the feasibility of on-the-fly correlation of network data and power consumption in data centers. During a preliminary 6-months long experimental campaign we created a dataset in a completely non-intrusive way with respect to the data center's network. The dataset allowed us to investigate the presence of correlation between power consumption and network traffic.
We found that, most of the time, power consumption and network traffic are strongly correlated (Pearson product-moment correlation coefficient greater than $0.7$). This allows to design novel approaches to improve resiliency of data centers that do not need to install any new software into the servers. Within the end of 2014 we count to complete the entire architecture and to deploy it in at least one MEF data center. 


As future work, we are investigating these advantages and we are considering fault injection techniques (\cite{arlat-laprie-96,Madeira,Cotroneo}) both on-line and off-line using  fault injection techniques \cite{protocolFaultInjection}). Studying effects of faults on network traffic and power consumption will help  to instrument failure prediction mechanisms. This will indeed allow to create a classification of correct and non-correct system behaviors that can recognize on-the fly when the system deviates from its normal behavior.


\section*{Acknowledgment}
This work was partially supported by the PRIN project TENACE:Protecting National Critical Infrastructures from Cyber Threats. The authors would like to thank the Italian Ministry of Economy and Finance for allowing the experimentation in their data centers.

\begin{small}
\addcontentsline{toc}{section}{References}
\bibliographystyle{unsrt}
\bibliography{main}
\end{small}

\end{document}